\documentclass[12pt]{article}
\input{epsf.tex}
\newlength{\dinwidth}
\newlength{\dinmargin}
\title {Nuclear Dynamics at the Balance Energy}
\author {Aman D. Sood and Rajeev K. Puri\\
\it Physics Department, Panjab University, Chandigarh -160 014, India\\}
\begin{document}
\maketitle
\begin{abstract}
We study the mass dependence of various quantities (like the average and
maximum density, collision rate, participant-spectator matter, temperature as
well as time zones for higher density) by simulating the reactions at the 
energy of vanishing flow.
This study is carried out within the framework of
Quantum Molecular Dynamics model. Our findings clearly indicate an existence
of a power law in all the above quantities calculated at the balance energy. The only
significant mass dependence was obtained for the temperature reached in the central
sphere. All other quantities are rather either insensitive or depend weakly on
the system size at balance energy. The time zone for higher
density as well as the time of maximal density and
collision rate follow a power law inverse to the energy of vanishing flow.
\end{abstract}
\section{Introduction}
\par It is now well established that the interactions at low
incident energies are dominated by the attractive part of the nuclear mean
field causing the emission of particles in the backward angles. 
These interactions, however, become repulsive at higher incident energies 
that pushes the particles in the forward (positive) angles. While going from the 
low incident energies to higher incident energies, there is a particular energy at which
the net flow disappears \cite {mol85}. This energy, (termed as
``balance energy'') $E_{bal}$ has been found to be of significant importance for the
understanding of the nature of nuclear interactions and related dynamics
[2-24].
\par Recently, the $E_{bal}$ was measured in $^{197}$Au+$^{197}$Au collisions
\cite{mag0061,mag0062}, extending the mass range of $E_{bal}$ between 24 and 394
units. In addition to $^{197}$Au+$^{197}$Au collisions,
one has also measured the $E_{bal}$ in
$^{12}$C+$^{12}$C \cite{west93}, $^{20}$Ne+$^{27}$Al \cite{west93},
$^{36}$Ar+$^{27}$Al \cite{ang97,buta95}, $^{40}$Ar+$^{27}$Al\cite{sull90},
$^{40}$Ar+$^{45}$Sc \cite{mag0062,west93,pak97},
$^{40}$Ar+$^{51}$V \cite{krof91}, $^{64}$Zn+$^{27}$Al \cite{he96},
$^{40}$Ar+$^{58}$Ni \cite{cus02}, $^{64}$Zn+$^{48}$Ti \cite{buta95},
$^{58}$Ni+$^{58}$Ni \cite{mag0062,cus02,pak97}, $^{58}$Fe+$^{58}$Fe \cite{pak97}, 
$^{64}$Zn+$^{58}$Ni \cite{buta95},
$^{86}$Kr+$^{93}$Nb \cite{mag0062,west93}, $^{93}$Nb+$^{93}$Nb \cite{krof92},
$^{129}$Xe+$^{118}$Sn \cite{cus02} and $^{139}$La+$^{139}$La \cite{krof92}
collisions. Most of the above reactions are symmetric and
central in nature. Some attempts are also reported in the
literature that deal with the impact parameter dependence of the balance
energy $E_{bal}$ \cite{mag0062,cus02,sull90,he96,pak97,xu92,soff95,suneel98}.
\par Interestingly, most of the theoretical attempts for disappearance of flow
are within the Boltzmann-Uehling-Uhlenbeck (BUU) model
\cite{mol85,krof92,mag0061,mag0062,west93,sull90,he96,pak97,li93,mota92,xu92,zhou94}.
Some attempts, however, are also made within the framework of Quantum
Molecular Dynamics (QMD) model
\cite {pak97,leh96,soff95,suneel98,soodprl,sood03,soodsym}. Note that
among all these attempts, only a few
deal with the mass dependence of the disappearance of
flow \cite {mag0061,mag0062,west93,zhou94,soodprl,sood03,soodsym}. There one reported a
power law behavior ($\propto$$A^{\tau}$) in the $E_{bal}$. For the first
time, a complete study of the mass dependence of balance
energy was presented by us where as many as sixteen systems, with
mass between 47 and 476 were analyzed \cite {soodprl,sood03,soodsym}.
The excellent agreement between the experimental measurements
and theoretical calculations
allowed us to predict the balance energy in
$^{238}$U+$^{238}$U collision around 37-39 MeV/nucleon \cite {soodprl}.
None of the above studies was extended to other heavy-ion phenomena 
at balance energy. Refs. \cite{he96,li93,zhou94,soff95,suneel98},
give some information about the nature of other variables at the
balance energy.
\par We here present a complete
analysis of the nuclear dynamics at balance energy by analyzing 
more than 14 (nearly symmetric) reactions as reported in ref.
\cite {sood03,soodsym}. Our present motivation is to investigate
whether other dynamical variables (apart from the disappearance of flow) show
a mass dependence at the balance energy or not. This present study is made within
the framework of Quantum Molecular Dynamics (QMD) model
\cite{leh96,soff95,suneel98,soodprl,sood03,soodsym,aich,hart98,pei89,singh00,khoanpa}.
Section II describes the
model in brief. Our results along with the discussion are presented
in section III. We summarize the results in section IV.\
\section{The Model}
The present study is made within the framework of the
Quantum Molecular Dynamics (QMD) model
\cite{suneel98,aich,hart98} where each nucleon interacts
via two- and three-body
interactions that preserve the nucleon-nucleon (nn) correlations and fluctuations.
Here each nucleon is represented by a Gaussian wave packet with width $\sqrt{L}$
centered around the mean position $\vec r_i(t)$ and the mean momentum
$\vec p_i(t)$:
\begin{equation}
\phi_i(\vec{r}, \vec{p}, t)=\frac{1}{(2\pi L)^{3/4}}e^{\left[-(\vec{r}-
\vec{r}_i (t))^2 /4L \right]}e^{\left[i{\vec p}_i (t)\cdot \vec{r}/\hbar
\right]}.
\label{e1}
\end{equation}
The Wigner distribution of a system with ($A_{T}$+$A_{P}$) nucleons is given by:
\begin{eqnarray}
f(\vec{r},\vec{p},t)& =& \sum_{i=1}^{A_T+A_P} \frac{1}{(\pi \hbar)^3}
e^{\left[-(\vec{r}-\vec{r}_i (t))^2/2L \right]}\times
\nonumber\\
&  & {e^{\left[-(\vec p-\vec p_i (t))^{2}2L/{\hbar}^2 \right]}.}
\end{eqnarray}
The nucleons propagate under the classical equations of motion:
\begin{equation}
\frac{d\vec r_i}{dt} = \frac{\partial \langle H \rangle }{\partial \vec p_i};
\end{equation}
\begin{equation}
\frac{d\vec p_i}{dt} = - \frac{\partial \langle H \rangle}{\partial \vec r_i}.
\end{equation}
The Hamiltonian $\langle H \rangle$ is given by:
\begin{eqnarray}
\langle H \rangle &=& \langle T \rangle + \langle V \rangle; \nonumber \\
&=& \sum_i \frac{\vec{p}_i^2}{2m_i} + \sum_i \sum_{i<j} V_{ij}^{total},
\end{eqnarray}
with
\begin{equation}
V_{ij}^{total} = V_{ij}^{local} + V_{ij}^{Yuk} + V_{ij}^{Coul}.
\end{equation}
Here $V_{ij}^{local}$, $V_{ij}^{Coul}$ and $V_{ij}^{Yuk}$ stand, respectively,
for the Skyrme, Coulomb and Yukawa parts of the nn interaction.
Following refs. \cite{soodprl,sood03,soodsym}, we use a hard equation of state
along with energy independent cross-section of 40 mb strength. This combination
is reported to explain the experimental balance energy nicely \cite{sood03}.
\section{Results and Discussion} As stated in ref. \cite {sood03},
a hard equation of state along with energy
independent nn cross-section of 40 mb strength yields a power law behavior 
$\propto$A$^{\tau}$. The power law ($c.A^{\tau}$) over the experimental points
yields $\tau=-0.42079\pm0.04594$, whereas our theoretical calculations with nn
cross-section of 40 mb strength had $\tau=-0.41540\pm0.08166$
\cite{sood03}. It is worth mentioning that this was the closest agreement
obtained so far. For the present mass dependent analysis,
we simulated the reactions of $^{20}$Ne+$^{27}$Al (b=2.6103 fm),
$^{36}$Ar+$^{27}$Al (b=2 fm), $^{40}$Ar+$^{27}$Al (b=1.6 fm),
$^{40}$Ar+$^{45}$Sc (b=3.187 fm), $^{40}$Ar+$^{51}$V (b=2.442 fm),
$^{40}$Ar+$^{58}$Ni (b=0-3 fm), $^{64}$Zn+$^{48}$Ti (b=2 fm),
$^{58}$Ni+$^{58}$Ni (b=2.48 fm), $^{64}$Zn+$^{58}$Ni (b=2 fm),
$^{86}$Kr+$^{93}$Nb (b=4.07 fm), $^{93}$Nb+$^{93}$Nb (b=3.104 fm),
$^{129}$Xe+$^{Nat}$Sn (b=0-3 fm), $^{139}$La+$^{139}$La (b=3.549 fm) and
$^{197}$Au+$^{197}$Au (b=2.5 fm) at their corresponding theoretical
balance energy$\footnote{The theoretical balance energy was calculated
by extrapolating the flow at two different energies with a step of $\pm$10
MeV/nucleon \cite{sood03}.}$ which is, respectively, 119, 74, 67.3, 89.4, 67.8, 64.6,
59.3, 62.6, 56.6, 69.2, 57, 49, 51.6, and 43 MeV/nucleon.
The reactions were followed till nuclear transverse flow saturates. As
noted from the above, the balance energy is smaller in heavier
colliding nuclei, compared to lighter one. As a result, one would expect
early saturation in lighter colliding nuclei compared to heavy one.
Though, the transverse flow saturates much early in lighter nuclei,
some of the quantities, however,
keep changing, therefore, we follow all the reactions uniformly till 300 fm/c. 
\par In the following,
we shall first study the time evolution and then shall present the mass dependence
of different quantities.
\par\subsection{\bf The Time Evolution} One of the motivations behind studying a
heavy-ion
collision is to extract the information regarding the hot and dense
nuclear matter. In our approach, the matter density is calculated by
\cite {khoanpa};
\begin{equation}
\rho(\vec{r},t)=\sum_{i=1}^{A_{T}+A_{P}}\frac{1}{(2\pi L)^{3/2}}
e^{(-(\vec{r}-\vec{r}_{i}(t))^{2}/2L)}.
\end{equation}
Here $A_{T}$ and $A_{P}$ stand, respectively, for the target and projectile.
In actual calculations,
we take a sphere of 2 fm radius around the center-of-mass and compute the
density at each time step during the reaction using eq. (7).
Naturally, one can either extract an average density $\langle \rho^{avg} \rangle$
over the whole sphere
or a maximal value of the density $\langle \rho^{max} \rangle$ reached anywhere in
the sphere. In fig. 1(a), we display the 
$\langle \rho^{avg} \rangle$/$\rho_{0}$
whereas fig. 1(b) shows the $\langle \rho^{max} \rangle$/$\rho_{0}$
as a function of the reaction time.
The displayed reactions are $^{20}$Ne+$^{27}$Al (A=47), $^{40}$Ar+$^{45}$Sc
(A=85), $^{64}$Zn+$^{58}$Ni (A=122), $^{93}$Nb+$^{93}$Nb (A=186),
$^{139}$La+$^{139}$La (A=278) and $^{197}$Au+$^{197}$Au
(A=394) spreading over the whole mass range. We see that the maximal
$\rho^{avg}$ is slightly higher for lighter systems compared to heavy ones.
A similar trend can also be seen for the evolution of $\rho^{max}$.
Further, the maximal value of the density for medium and heavy systems
is comparable with the average one. This clearly indicates
that the dense matter is formed widely and uniformly in the central region
of 2 fm radius. On the other hand, substantial difference in two densities
for lighter colliding nuclei indicate that the dense matter is not
homogenous and uniform in these reactions. Due to high incident energy,
$^{20}$Ne+$^{27}$Al reaction
finishes much early compared to $^{197}$Au+$^{197}$Au which is simulated at
a relatively lower incident energy. 
Similarly, 
the peaks in (the maximum $\langle \rho^{max} \rangle$ and average
$\langle \rho^{avg} \rangle$) densities are
also delayed in heavier nuclei compared to lighter one. The spreading of the
high density zone in heavier colliding nuclei over the long time span indicates
the on going interactions among nucleons.
This result is in agreement with \cite{khoanpa}.
\par Another quantity
directly linked with the density is the collision rate. In fig. 2, we
display the $dN_{coll}/dt$ versus time. This rate represents the net 
collisions after fulfilling the Pauli principle. Naturally, the attempted rate
will be much higher than the allowed one.
Due to larger interaction volume, the interactions among nucleons
in heavy nuclei continue for a long time.
This effect should be obvious if one looks the density
profile (see fig. 1). A finite density zone will naturally lead to more and
more nn collisions
and as a result, the collision rate will be more for heavy colliding nuclei.
\par As stated above, all the reactions are simulated at the balance energy where
attractive forces counter balance the repulsive forces. This fact should also be
reflected in the quantities like the spectator and participant matter.
All nucleons having experienced at least one collision are
counted as {\it participant matter}. The remaining matter is labeled as
the {\it spectator matter}. The nucleons with more than one collision are labeled as
{\it super-participant matter}. This definition gives us possibility to analyze the
reaction in terms of participant-spectator fireball model.
\par In fig. 3, we display
the normalized spectator matter (upper part) and participant matter (lower part)
as a function of the time. At the start of the reaction, all nucleons constitute
spectator matter. Therefore, no participant matter exists at
t=0 fm/c. Since the $^{20}$Ne+$^{27}$Al reaction happens at a
relative higher energy
(=119 MeV/nucleon), the transition from the
spectator to participant matter is swift and fast.
On the other hand, due to low bombarding energy,
the transition from the spectator to
participant matter in heavier colliding nuclei is rather slow and gradual.
Interestingly, at the end,
all reactions (that happen between incident energy 43 and 119
MeV/nucleon) lead to nearly same participant matter indicating the
universality in the balancing of attractive and repulsive forces.
\par From the above facts, it is clear that heavier colliding nuclei (at
$E_{bal}$) have delayed and expanded evolution of the density and
participant-spectator
matter. It will be of more interest to see how their mass dependence
behaves like. This will be discussed in the following paragraphs.
\par\subsection {\bf The Mass Dependence} In fig. 4, we display the maximal
value of $\langle \rho^{avg} \rangle$
and $\langle \rho^{max} \rangle$ versus composite mass of the system.
Note that all the reactions, considered here, are symmetric i.e.
$\eta~(\begin{array}{c}\left|
{\frac {A_{T}-A_{P}}{A_{T}+A_{P}}}\right|\end{array})<0.2$.
Very interesting, we see
that the maximal value follows a power law $\propto$$A^{\tau}$ with $\tau$
being $-0.05182\pm0.00776$ for average density $\langle \rho^{avg} \rangle$
and $-0.11477\pm0.01217$ for maximum density
$\langle \rho^{max} \rangle$.$\footnote{A small deviation can be seen in
the cases where $\eta\ne{0}$.}$ 
In other words, a slight decrease in the density occurs with the size of
the system. This decrease is much smaller compared to the $E_{bal}$
($\tau_{expt}=-0.42079\pm0.04594$ and $\tau_{th}=-0.41540\pm0.08166$).
Had these reactions being
simulated at a fixed incident energy, the trend would
have been totally different \cite{khoanpa}. Since lighter nuclei cannot
be compressed easily, their maximal density at a fixed incident
energy will be less compared to heavy nuclei. 
Since $E_{bal}$ in the present case scales as $A^{\tau}$, a weak mass dependence
is also observed in density profiles. 
\par The mass dependence of the (allowed) nn collisions is depicted
in fig. 5. Here one sees a (nearly) linear enhancement in the nn
collisions with the
size of the interacting system. This enhancement can be explained mainly using
a power law $\propto$$A^{\tau}$; with $\tau=0.87829\pm0.01833$. If 
one keeps the incident energy fixed (e.g. in the figure, we kept E=50 MeV/nucleon),
the nn collisions should scale as ``A''. Our fitting gives $\tau=1.04633\pm0.01712$,
which is very close to unity.
\par A dynamical quantity that can serve as an indicator of the role of repulsive and
attractive forces is the participant and spectator matter. Naturally, the
possibility of a collision will depend upon the mean free path of nucleons.
Similar is the case of spectator and participant matter. In fig. 6, we
display the spectator, participant and super-participant matter (obtained at
300 fm/c) as
a function of the total mass of the system. Very interesting, we see a nearly
mass independent behavior of the participant matter
($\tau=-0.03621\pm0.00954$).
Similar behavior occurs in the case of spectator matter ($\tau=0.08323\pm0.02232$). 
A slight deviation occurs in the case of
$^{20}$Ne+$^{27}$Al. Some small fluctuations can also be due to the variation in
the impact parameter, which is not fixed in the present study. The choice of
impact parameter is 
guided by the experimental measurements.
As noted in ref. \cite {suneel98}, the variation in the impact parameter can have
drastic influence on the participant/spectator matter.
The super-participant matter shows a little more dependence on the mass of the
system ($\tau=-0.14241\pm0.04184$). This can be understood again by looking at
the density profile (fig. 4).
There one concluded that the lighter nuclei lead to higher densities. In other
words, the mean free path will be smaller in lighter nuclei that results
in more nn collisions. One should keep in the mind that the mass
independent nature of the participant matter is not a trivial
observation. For a fixed
system mass, the participant matter depends linearly
on the incident energy. In the present case, though the mass of the system
increases, their corresponding energy decreases, resulting in the net mass
independent nature. 
This also suggests that the repulsive and attractive forces at $E_{bal}$
counter-balance each other in such a manner that the net
participant matter remains the same. One may also say that since the contribution
of the mean field towards transverse flow is nearly mass independent
\cite{sood03,blat91}, one needs same amount of participant matter to counter
balance the attractive forces. In other words, the participant matter can
act as a barometer to study the balance energy in heavy-ion collisions.
\par The associated quantity linked with the dense matter is the temperature. In
principle, a true temperature can be defined only for a thermalized and
equilibrated matter. Since in a heavy-ion collision, the matter is non-equilibrated,
one can not talk of ``temperature''. One can, however, look in
terms of a local environment only. In our present
case, we follow the description of the temperature given in
refs. \cite{khoanpa,purinpa,khoanpa1}.
Several different authors have given different descriptions
of the local or global temperatures \cite{bert88,stoc86,neise90,lang90}.
Some studies of temperature are based on the fireball model \cite{bert88},
whereas others take the hydrodynamical model into account \cite{stoc86}. In refs.
\cite{stoc86,neise90}, the thermalization is directly connected with the
non-diagonal elements of the stress tensor. One has even defined the
``transverse'' temperature in terms of $\langle P_{I}^{2}/2m \rangle$;
$P_{I}^{2}$ is the average transverse momentum squared \cite{stoc86}. In the
present case, the extraction of the temperature ``T'' is based on the local density
approximation i.e. one deduces the temperature in a volume element surrounding
the position of each particle at a given time step \cite {khoanpa,purinpa,khoanpa1}.
Here, we postulate that each local volume element of nuclear matter in coordinate
space and time has some ``temperature'' defined by the diffused edge of the deformed
Fermi distribution consisting of two colliding Fermi spheres which is typical for
a non-equilibrium momentum distribution in heavy-ion collisions.
\par In this formalism, (dubbed as hot Thomas-Fermi approach \cite{khoanpa}),
one determines the
extensive quantities like the density, kinetic energy, as well as the 
entropy with the help of
momentum distributions at a given temperature. For more details, reader is
referred to refs. \cite{khoanpa,purinpa,khoanpa1}. Using this formalism, we also
extracted the average and maximum temperature within a central sphere of 2 fm
radius as described in the case of density.
\par In fig. 7, we plot the maximal
value of $\langle T^{avg} \rangle$ and $\langle T^{max} \rangle$
as a function of the composite mass of the system.
Some fluctuations can be seen in these plots that can be due to the choice of
impact parameter as well as incident energy \cite{khoanpa,khoanpa1,stoc86,puri94,purihirschegg}.
As stated above, the impact parameter choice is guided by the experimental
constraints. Further, the $E_{bal}$ was extracted using a straight line
interpolation, therefore, both
these factors may add to the present fluctuations. 
One sees that both quantities can be parameterized in terms of a
power law function $\propto$$A^{\tau}$; The power factor
$\tau$ is rather quite large (being equal to $-0.83743\pm0.11355$ and
$-0.51079\pm0.08218$), respectively, for the average and maximal temperature.
This sharp mass dependence in the temperature is rather in
contradiction to the mild mass dependence obtained in all other quantities.
This is not astonishing since
temperature depends, crucially, on the kinetic energy
(or the excitation energy) of the system \cite{khoanpa,purinpa,khoanpa1,puri94,purihirschegg}.
It was shown in ref. \cite{puri94,purihirschegg} that
for a given colliding geometry, the maximal value of the
temperature does not depend
upon the size of the interacting source. Rather it depends only on the
bombarding energy.
\par In fig. 8, we display the time of maximal collision rate and
average density. We see a power law behavior in both the quantities. 
The small balance energy in heavy nuclei delays the maximal compression. 
Interestingly, the power factor $\tau$ is close to (1/3) in both the cases.
The
$E_{bal}$ was shown to scale with power factor $\tau\approx{-0.4}$.
In other words, the time of
maximal collision rate and density
varies approximately as a inverse of the $E_{bal}$.
\par Apart from the maximal quantities,
another interesting quantity is the dense zone at the balance energy.
This is depicted in fig. 9 where we display the time interval for which
$\rho$$\geq$$\rho_{0}$ (upper part) and $\rho$$\geq$$\rho_{0}/2$. Again
both quantities follow a
power law behaviour. Interestingly, the time intervals for the high density
have a power law dependence with $\tau=0.33853\pm0.05358$ and
$\tau=0.46833\pm0.04265$, respectively, for $\rho$$\geq$$\rho_{0}$ and
$\rho$$\geq$$\rho_{0}/2$ which is again very
close to the inverse of the mass dependence of $E_{bal}$. 
This also points toward the fact that the formation and identification
of the fragments is delayed in heavier nuclei compared to lighter nuclei. This
conclusion is in agreement with earlier
calculations \cite {singh00,purihirschegg}.
\section {Summary}
\par Using the QMD model, we presented the mass dependence of
various quantities (such as average and maximum central density, temperature,
collision dynamics, participant and spectator matter as well as the time zone
for hot and dense nuclear matter) at the energy of vanishing flow ($E_{bal}$).
This study was conducted using a hard equation of state along with nn
cross-section of 40 mb strength. This combination is reported to explain the 
experimentally extracted balance energy for large number of
cases \cite{sood03}.
Our calculations present several interesting facts:
\par The reaction saturation time is smaller for lighter nuclei compared
to heavy nuclei. The maximal values of density, temperature and collision rate is
also shifted accordingly. In all the cases (i.e. in average and maximum central density,
temperature, participant and spectator matter etc.), a power law dependence can
be seen. The only quantity where power factor $\tau$ is significant
(with $\tau$$\geq$$|0.2|$) is the temperature reached in the central zone.
Other quantities are nearly mass independent.
The mass independent nature of the participant matter makes it a good
alternate indicator for determining the balance energy. The existence of
dense zone scales as inverse of the energy of vanishing flow.\\

{\it This work is supported by the grant (No. SP/S2/K-21/96)
from Department of Science and Technology, Government of India.}

\newpage
{\large \bf Figure Captions:}\\

{\bf Fig.  1:} The evolution of (a) average density $\langle \rho^{avg} \rangle$ and
(b) the maximum density $\langle \rho^{max} \rangle$
reached in a central sphere of radius 2 fm as a function of time. Here
reactions of $^{20}$Ne+$^{27}$Al (b=2.6103 fm), $^{40}$Ar+$^{45}$Sc
(b=3.187 fm), $^{64}$Zn+$^{58}$Ni (b=2 fm),
$^{93}$Nb+$^{93}$Nb (b=3.104 fm), $^{139}$La+$^{139}$La (b=3.549 fm)
and $^{197}$Au+$^{197}$Au (b=2.5 fm) are simulated at their corresponding
theoretical balance energies (for details, see the text). The shaded area
represents the reaction of $^{20}$Ne+$^{27}$Al.\\

{\bf Fig.  2:} Same as fig. 1(a), but the rate of allowed collisions
$dN_{coll}/dt$ versus reaction time.\\

{\bf Fig.  3:} Same as fig 1, but the time evolution of normalized
spectator matter (upper part) and participant matter (lower part).\\ 

{\bf Fig.  4:} The maximal value of the average density
$\langle \rho^{avg} \rangle_{max}$ (upper part) and maximum density
$\langle \rho^{max} \rangle_{max}$ (lower part) as a function
of the composite mass of the system. The solid lines are the fits
to the calculated
results using $c.A^{\tau}$ obtained with $\chi^{2}$ minimization.\\

{\bf Fig.  5:} The total number of the allowed collisions (obtained at the final stage)
versus composite mass of the system. The solid squares and open diamonds are the
results obtained at $E_{bal}$ and 50 MeV/nucleon, respectively. The solid
and dashed lines are the fits obtained with procedure explained in fig. 4.\\

{\bf Fig.  6:} Same as fig. 5, but for the final saturated participant,
spectator and super-participant matter.\\

{\bf Fig.  7:} Same as fig. 5, but for the maximal value of the average temperature
(upper part) and maximum temperature (lower part).\\

{\bf Fig.  8:} Same as fig. 5, but for the time of maximal value of collision
rate (open stars) and average density (solid squares). The solid and
dashed lines represent the $\chi^{2}$ fits with power law $c.A^{\tau}$.\\

{\bf Fig.  9:} Same as fig. 8, but for the time zone for
($\rho$$\geq$$\rho_{0}$) (upper part) and for ($\rho$$\geq$$\rho_{0}/2$)
(lower part) as a function of composite mass of the system.             

\end{document}